\documentclass[final,3p,nofootinbib,hyperref]{elsarticle}

\usepackage{amssymb}
\usepackage{amsmath}
\usepackage{graphicx}
\usepackage{changes}
\usepackage[mathlines]{lineno}
\usepackage{xspace}
\usepackage{url}
\usepackage{mediabb}
\usepackage{epstopdf}
\usepackage[normalem]{ulem}
\usepackage[varg]{txfonts} 
\usepackage{graphicx,color}
\usepackage{amsmath,amssymb,amsfonts}
\usepackage[colorlinks=true,linkcolor=blue,plainpages=false,hypertex]{hyperref}

\def \be {\begin{equation}}
\def \ee {\end{equation}}
\def \ee  {\end{equation}}
\def \bea {\begin{eqnarray}}
\def \eea {\end{eqnarray}}

\newcommand{\vone}{$\mathrm{v_{1}\,}$}
\newcommand{\pT} {\mathrm{p_{T}}}

\def \lt {\mbox{$<$}}

\journal{Phys. Lett. B}

\begin{document}
\begin{frontmatter}
%


\title{Beam energy dependence of rapidity-even dipolar flow in Au+Au collisions}



\author{
J.~Adam$^{9}$,
L.~Adamczyk$^{1}$,
J.~R.~Adams$^{31}$,
J.~K.~Adkins$^{21}$,
G.~Agakishiev$^{19}$,
M.~M.~Aggarwal$^{33}$,
Z.~Ahammed$^{56}$,
N.~N.~Ajitanand$^{44}$,
I.~Alekseev$^{17,28}$,
D.~M.~Anderson$^{46}$
% ...
\\
(STAR Collaboration)
}
\address{$^{1}$AGH University of Science and Technology, FPACS, Cracow 30-059, Poland}
% ...
\address{$^{9}$Creighton University, Omaha, Nebraska 68178}
% ...
\address{$^{17}$Alikhanov Institute for Theoretical and Experimental Physics, Moscow 117218, Russia}
% ...
\address{$^{19}$Joint Institute for Nuclear Research, Dubna, 141 980, Russia}
% ...
\address{$^{21}$University of Kentucky, Lexington, Kentucky 40506-0055}
% ...
\address{$^{28}$National Research Nuclear University MEPhI, Moscow 115409, Russia}
% ...
\address{$^{31}$Ohio State University, Columbus, Ohio 43210}
% ...
\address{$^{33}$Panjab University, Chandigarh 160014, India}
% ...
\address{$^{44}$State University of New York, Stony Brook, New York 11794}
% ...
\address{$^{46}$Texas A\&M University, College Station, Texas 77843}
% ...
\address{$^{56}$Variable Energy Cyclotron Centre, Kolkata 700064, India}
% ...


\begin{abstract}
New measurements of directed flow for charged hadrons, characterized by the 
Fourier coefficient \vone,  are presented for  transverse momenta $\pT$, 
and centrality intervals in Au+Au collisions recorded by the STAR experiment for 
the center-of-mass energy range $\mathrm{\sqrt{s_{_{NN}}}} = 7.7 - 200$ GeV. 
The measurements underscore the importance of momentum conservation, and the 
characteristic dependencies on $\mathrm{\sqrt{s_{_{NN}}}}$, centrality and $\pT$ are 
consistent with the expectations of geometric fluctuations generated in the initial stages 
of the collision, acting in concert with a hydrodynamic-like expansion.
The centrality and $\pT$ dependencies of $\mathrm{v^{even}_{1}}$, as well as  
an observed similarity between its excitation function and that for $\mathrm{v_3}$,  
could serve as constraints for initial-state models. The $\mathrm{v^{even}_{1}}$  
excitation function could also provide an important supplement to the flow measurements 
employed for precision extraction of the temperature dependence of the 
specific shear viscosity.
\end{abstract}


\end{frontmatter}


\begin{twocolumn}

High-energy nuclear collisions at the Relativistic Heavy Ion Collider (RHIC) and the Large Hadron Collider (LHC) 
can result in the creation of a plasma composed of strongly coupled quarks and gluons (QGP). 
Full characterization of this hot and dense matter is a major goal of present-day high-energy physics research.  
Recent studies have emphasized the use of anisotropic flow measurements
to study the transport properties of this matter 
\cite{Teaney:2003kp,Lacey:2006pn,Romatschke:2007mq,Schenke:2011tv,Song:2011qa,Shen:2010uy,Gardim:2012yp,Niemi:2012ry,Qin:2010pf}. 
A current focus is centered on delineating the role of  initial-state fluctuations, as well as reducing their influence 
on the uncertainties associated with the extraction of the temperature dependent specific shear 
viscosity (i.e. the ratio of shear viscosity to entropy density $\mathrm{\frac{\eta}{s}(T)}$) of the QGP produced
in these collisions~\cite{Schenke:2011tv,Song:2011qa,Shen:2010uy,Gardim:2012yp,Niemi:2012ry,Qin:2010pf,
Alver:2010gr,Schenke:2010rr,Lacey:2013eia,McDonald:2016vlt,Bernhard:2016tnd}. 

%
%
 \begin{figure*}[t]
\centering
\includegraphics[width=.95\textwidth,angle=0]{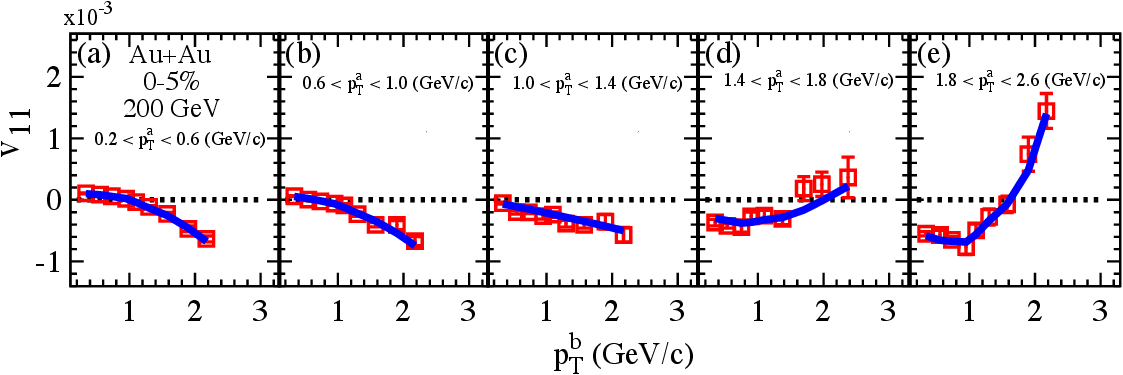}
\vskip -0.4cm
\caption{$\mathrm{v_{11}}$ vs. $p_{T}^{\text{b}}$ for several selections of $p_{T}^{\text{a}}$  for 
0-5\% central Au+Au collisions at $\mathrm{\sqrt{s_{_{NN}}}} = 200$~GeV. The curve shows the 
result of the simultaneous fit with Eq.~(\ref{corrv1}). The fit resulted 
in the value $\chi^2 = 1.1$ per degree of freedom  (see text). 
 \label{v11f}
 }
\end{figure*}

The $\mathrm{v_n}$ coefficients used to characterize anisotropic flow, are normally obtained from 
a Fourier expansion of the azimuthal angle ($\mathrm{\phi}$) distribution of the particles produced orthogonal 
to the beam direction \cite{Ollitrault:2012cm,Poskanzer:1998yz}:
\begin{eqnarray}
\label{eq:1}
\mathrm{\frac{dN}{d\phi}\propto1+2\sum_{n=1}^{\infty} v_{n}\cos n(\phi-\Psi_{n})},
\end{eqnarray}
where $\mathrm{\Psi_n}$ represents the $\mathrm{n^{th}}$ order event plane, 
i.e., $\mathrm{\langle e^{in\phi}\rangle} = \mathrm{v_n e^{in\Psi_{n}}}$ and the brackets 
indicate averaging over particles and events. The coefficient \vone is commonly 
termed directed flow, $\mathrm{v_2}$ is the elliptic flow, $\mathrm{v_3}$ is the triangular flow etc. 
For flow dominated distributions, the $\mathrm{v_n}$ coefficients are related to the 
Fourier coefficients $\mathrm{v_{nn}}$ used to characterize two-particle correlations 
in relative azimuthal angle $\mathrm{\Delta\phi=\phi_{\mathrm{a}}-\phi_{\mathrm{b}}}$ for 
particle pairs $\mathrm{a,b}$ \cite{Lacey:2005qq}:
\begin{eqnarray}
\label{eq:2}
\mathrm{\frac{dN^{pairs}}{d\Delta\phi}\propto1+2\sum_{n=1}^{\infty}\mathrm{v_{nn}}\cos (n \Delta\phi)}.
\end{eqnarray}
However, so-called  non-flow (NF) correlations can also contribute to the two-particle 
correlations~\cite{Lacey:2005qq,Borghini:2000cm,Luzum:2010fb,Retinskaya:2012ky,ATLAS:2012at}:
\begin{eqnarray}
\label{eq:3}
\mathrm{v_{nn}}(\pT^{\text{a}},\pT^{\text{b}})  = \mathrm{v_n}(\pT^{\text{a}})\mathrm{v_n}(\pT^{\text{b}})+ \delta_{\text{NF}},
\end{eqnarray}
where $\delta_{\text{NF}}$ includes possible contributions from resonance decays, Bose-Einstein correlations, jets, and global momentum conservation (GMC).
%
%
 \begin{figure*}[t]
\centering{
\includegraphics[width=0.80\linewidth,angle=0]{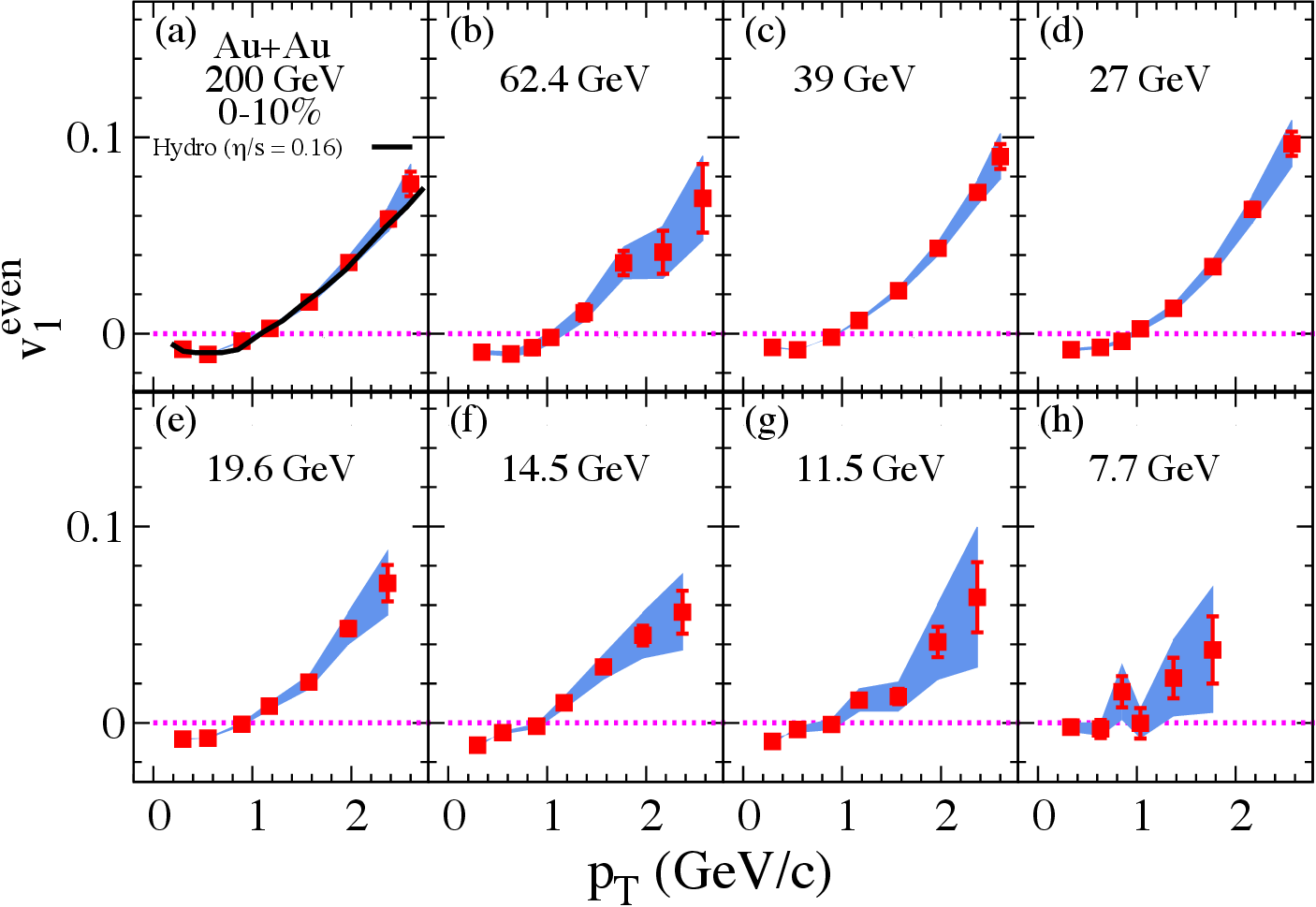}
\vskip -0.4cm
\caption{ Extracted values of $\mathrm{v^{even}_{1}}$ vs. $\mathrm{p_{T}}$ for 0-10\% central Au+Au collisions for several values of  $\mathrm{\sqrt{s_{_{NN}}}}$ as indicated; the $\mathrm{v^{even}_{1}}$ values are obtained via fits with Eq.~(\ref{corrv1}). The curve in panel (a) shows the result from a hydrodynamic calculations \cite{Retinskaya:2012ky}. The shaded bands indicate the systematic uncertainties.
 \label{v1_pt}
 }
}
\end{figure*}
%
%
 \begin{figure*}
\centering{
\includegraphics[width=6cm,height=8.0cm,angle=-90]{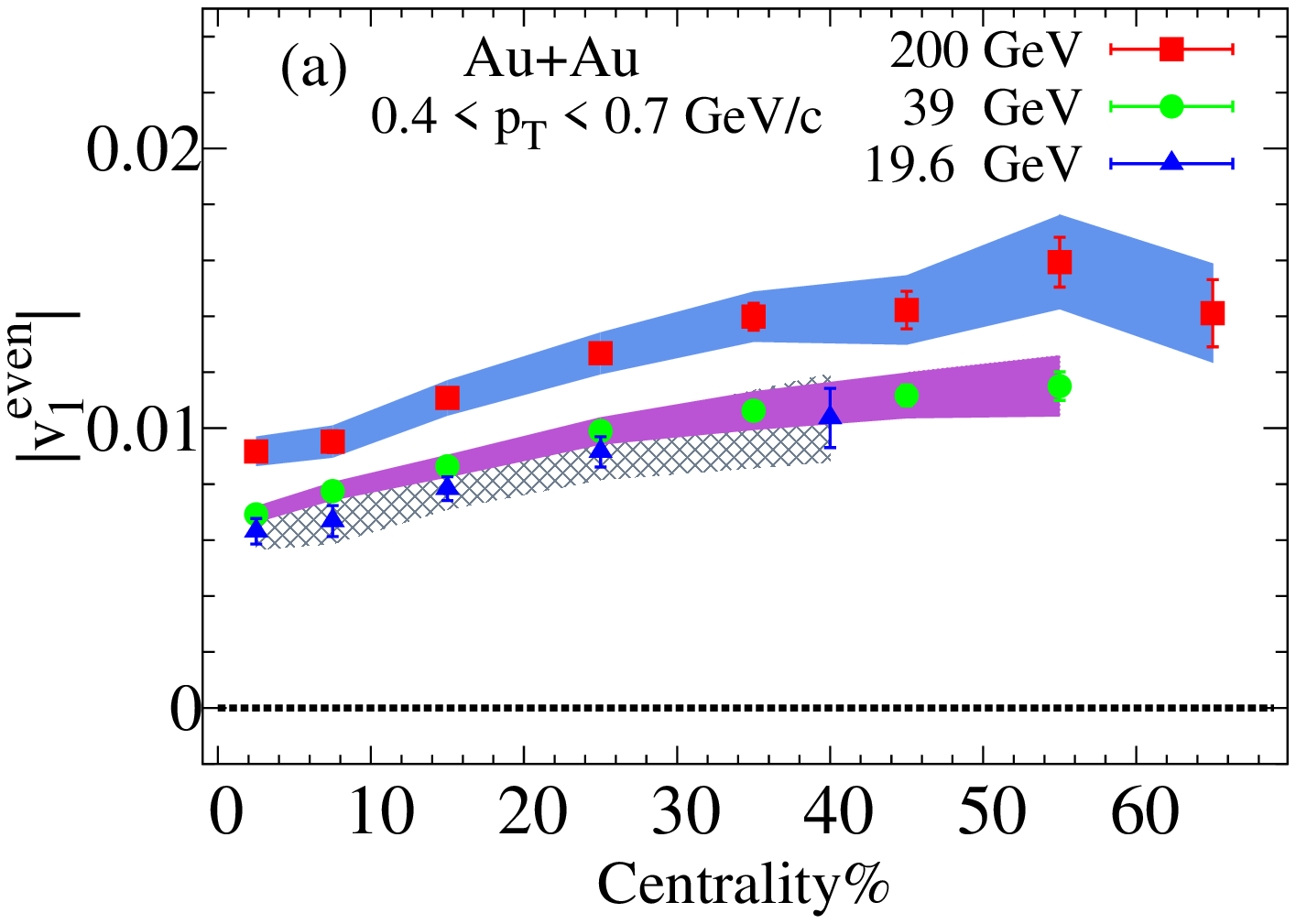}\hspace{-1.6cm}
\includegraphics[width=6.35cm,height=9.6cm ,angle=-90]{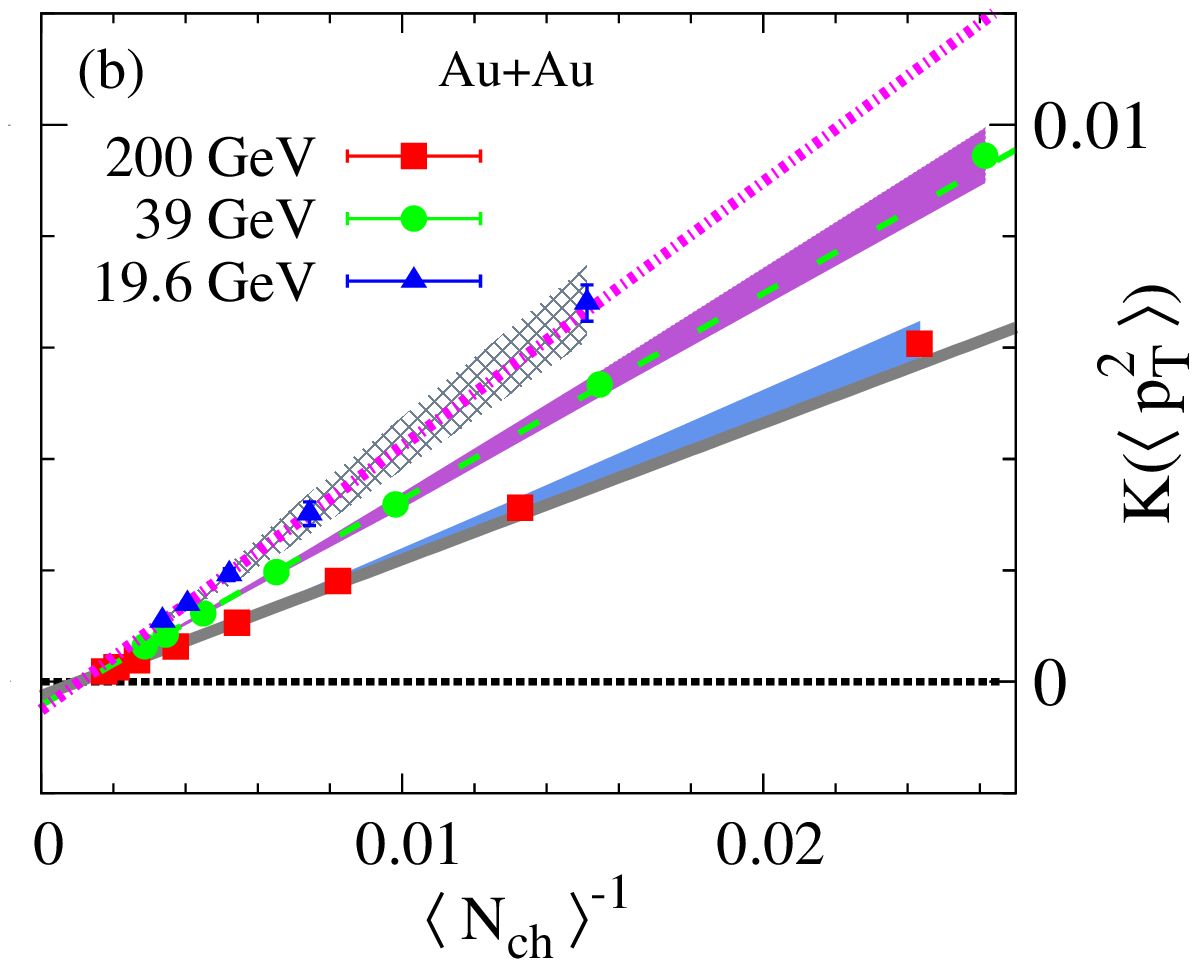}
\vskip -0.5cm
\caption{(a) Centrality  dependence of $\mathrm{v^{even}_{1}}$ for $0.4 \lt \pT \lt 0.7$~GeV/c 
for Au+Au collisions at $\mathrm{\sqrt{s_{_{NN}}}} = 200, 39,$ and $19.6$~GeV; 
(b) $\mathrm{K}$ vs. $\mathrm{\langle N_{ch} \rangle^{-1}}$ for the $\mathrm{v^{even}_{1}}$ values 
shown in (a). The $\mathrm{\langle N_{ch} \rangle}$ values correspond to the centrality intervals 
indicated in panel (a). The $\mathrm{v^{even}_{1}}$ and $\mathrm{K}$ values are obtained via fits 
with Eq.~(\ref{corrv1}) (see text). The indicated lines show linear fits to the data; the 
shaded bands represent the systematic uncertainties.
 \label{cent_dep}
 }
}
\end{figure*}

In the absence of fluctuations, the directed flow $\mathrm{v_{1}}$ develops along 
the direction of the impact parameter \cite{Danielewicz:2002pu} and is an odd function, 
$\mathrm{v^{odd}_1(\eta)} = \mathrm{-v^{odd}_1(-\eta)}$, of pseudorapidity. However, initial-state 
fluctuations, acting in concert with hydrodynamic-like expansion, gives an additional  
rapidity-even, $\mathrm{v^{even}_1(\eta) = v^{even}_1(-\eta)}$, 
component \cite{Luzum:2010fb,Teaney:2010vd} resulting in the total:
\[
    \mathrm{v_1(\eta) = v^{even}_1(\eta) + v^{odd}_1(\eta)}.
\]
The magnitude of $\mathrm{v^{odd}_1(\eta)}$ can be made negligible via a symmetric pseudorapidity selection, to give a straightforward measurement of 
$\mathrm{v^{even}_1(\eta)}$.

The rapidity-even \vone is proportional to the fluctuations-driven  dipole 
asymmetry $\varepsilon_1$ of the system \cite{Luzum:2010fb,Teaney:2010vd,Gardim:2011qn}; 
$\mathrm{v^{even}_{1} \propto \varepsilon_1}$, 
where $\varepsilon_1 \equiv  \left< \lvert  r^3e^{i\phi} \rvert \right>/\left<r^3\right>$ 
and averaging is taken over the initial energy density after re-centering the coordinate system, 
i.e., $\left< \lvert r^3e^{i\phi} \rvert \right> = 0$.
Hydrodynamical model calculations \cite{Retinskaya:2012ky} indicate that the magnitude 
of $\mathrm{v^{even}_{1}}$ is sensitive to $\mathrm{\eta/s}$, albeit with less sensitivity than for the 
higher order harmonics, $\mathrm{n \ge 2}$. It has not been experimentally established whether 
this sensitivity depends on the temperature $\mathrm{T}$, baryon chemical potential $\mu_{B}$ or both. Similarly
is has not been established whether this sensitivity could reflect the influence of a possible 
critical end point (CEP) in the phase diagram for nuclear matter \cite{Lacey:2014wqa}. 
Therefore, differential $\mathrm{v^{even}_{1}}$ measurements that span a broad range of $\mathrm{\sqrt{s_{_{NN}}}}$ ($T$ and $\mu_{B}$), could potentially provide (i) unique supplemental constraints to discern between different initial-state models, (ii) aid precision extraction of $\eta/s$ and study its possible dependence on $\mathrm{T}$ and $\mathrm{\mu_B}$, and (iii) give insight on the CEP.  It is noteworthy that the paucity of $\mathrm{v^{even}_{1}}$ measurements at RHIC energies precludes their current use as constraints.

The present work  employs two-particle correlation functions to extract $\mathrm{v_{11}=\langle\cos \Delta\phi \rangle}$ values as a function of $\pT^{\text {a}}$, $\pT^{\text{b}}$ and centrality for a broad selection of beam energies. 
In turn the GMC ansatz~\cite{Borghini:2000cm,Borghini:2002mv} is used in conjunction with the two-component fitting procedure outlined in Refs.~\cite{Retinskaya:2012ky,ATLAS:2012at} and discussed below, to extract $\mathrm{v^{even}_{1}}$ as a function of $\pT$ and centrality for each value of $\mathrm{\sqrt{s_{_{NN}}}}$. 
The measurements indicate the characteristic $\pT$-dependent directed flow patterns associated with rapidity-even dipolar flow~\cite{Luzum:2010fb,Teaney:2010vd,Gardim:2011qn}, as well as striking centrality and $\mathrm{\sqrt{s_{_{NN}}}}$ dependencies which could  serve as constraints for initial- and final-state model inputs.

The data reported in this analysis are from Au+Au collisions spanning the full range of 
energies, $\mathrm{\sqrt{s_{_{NN}}}} = 7.7 - 200$~GeV, in beam energy scan I $(\text{BES-I})$, collected with the 
STAR detector using a minimum bias trigger.  The collision vertices were reconstructed using
charged-particle tracks measured in the Time Projection Chamber (TPC)~\cite{Anderson:2003ur}. 
The TPC covers the full azimuth and has a pseudorapidity range of $\mathrm{|\eta|}<1.0$.  
Events were selected to have a vertex position about the nominal center of the TPC (in the beam direction) 
of $\pm$ 30 cm at $\mathrm{\sqrt{s_{_{NN}}}} = 200$~GeV, $\pm$ 40 cm at $\mathrm{\sqrt{s_{_{NN}}}} = 62,~39,~27,~19.6$ and $14.5$~GeV, 
$\pm$ 50 cm at $\mathrm{\sqrt{s_{_{NN}}}} = 11.5$~GeV and $\pm$ 70 cm at $\mathrm{\sqrt{s_{_{NN}}}} = 7.7$~GeV, and 
to be within a radius of $1-2$~cm with respect to the beam axis.  
Note that the distribution of the vertex positions broadens (in the beam direction) as 
the beam energy is lowered.

The centrality of each collision was determined by measuring event-by-event multiplicity and interpreting 
the measurement with a tuned Monte Carlo Glauber calculation~\cite{Adamczyk:2012ku,Abelev:2009bw}.   
Analyzed tracks were required to have a distance of closest approach to the primary vertex to be less than
3~cm, and to have at least 15 TPC space points used in their reconstruction. Furthermore, the ratio of the number 
of fit points to the maximum possible number of TPC space points was required to be larger
than~0.52 to remove split tracks.  The $\pT$ of tracks was limited to the range $0.2<\pT<4$~GeV/$c$.

The correlation function technique~\cite{Lacey:2005qq} was used to generate the two-particle $\Delta\phi$ correlations,
\begin{eqnarray}\label{corr_func}
 C_r(\Delta\phi, \Delta\eta) = \frac{(dN/d\Delta\phi)_{\text{same}}}{(dN/d\Delta\phi)_{\text{mixed}}},
\end{eqnarray} 
where $\mathrm{\Delta\eta = \eta_{a} - \eta_{b}}$ is the pseudorapidity separation between the particle pairs $\mathrm{a,b}$, $\mathrm{(dN/d\Delta\phi)_{\text{same}}}$ represents the normalized azimuthal distribution of  particle pairs  from the same event and $\mathrm{(dN/d\Delta\phi)_{\text{mixed}}}$ represents the normalized azimuthal distribution for particle pairs in which each member is selected  from different events but with a similar classification for the vertex, and centrality. The pseudorapidity requirement  $\mathrm{|\Delta\eta| > 0.7}$ was also imposed on track pairs to minimize possible non-flow contributions associated with the short-range correlations from resonance decays, Bose-Einstein correlations and jets.

The two-particle Fourier coefficients $\mathrm{v_{nn}}$ are obtained from the correlation function as:
\begin{eqnarray}\label{vn}
 \mathrm{v_{nn}} &=& \frac{\sum_{\Delta\phi} C_r(\Delta\phi)\cos(n \Delta\phi)}{\sum_{\Delta\phi}~C_r(\Delta\phi)},
\end{eqnarray}
where the $\mathrm{\Delta\phi}$ bin width was chosen to optimize statistical significance. The $\mathrm{v_{nn}}$
values were then used to extract $\mathrm{v^{even}_{1}}$ via a simultaneous fit of $\mathrm{v_{11}}$ as 
a function of $\pT^{\text {b}}$ for several selections of  $\pT^{\text{a}}$ with Eq.~(\ref{eq:3}),
\begin{equation}
\label{corrv1}
\mathrm{v_{11}(\pT^{a},\pT^{b})  = v^{even}_{1}(\pT^{a})v^{even}_{1}(\pT^{b}) - K \pT^{a}\pT^{b}}.
\end{equation} 

 Here, $\mathrm{K \propto 1/(\langle N_{ch} \rangle \langle p_{T}^{2}\rangle)}$ takes into account  
the non-flow correlations induced by global momentum conservation~\cite{Retinskaya:2012ky,ATLAS:2012at}; 
$\mathrm{\langle N_{ch} \rangle}$ is the mean multiplicity and $\mathrm{\langle p_{T}^{2}\rangle}$ 
is  proportional to the variance of the transverse momentum over the full phase space. 
The charged particle multiplicity measured in the TPC acceptance is used as 
a proxy for $\mathrm{\langle N_{ch} \rangle}$.
For  a given centrality selection, the left hand side of  Eq.~(\ref{corrv1}) represents a $\text{N}$-by-$\text{M}$ $\mathrm{v_{11}}$ matrix  (i.e., $\text{N}$ values for $\pT^{b}$ for each of the $\text{M}$ $\pT^{a}$ selections) which we fit with the right hand side of Eq.~(\ref{corrv1}) using $\text{N} + 1$ parameters: N values of $\mathrm{v^{even}_{1}(\pT)}$ and one additional parameter $\mathrm{K}$, the coefficient of momentum conservation \cite{Jia:2012gu}.  Figure \ref{v11f} illustrates the efficacy of the fitting procedure for 0-5$\%$ central Au+Au collisions at $\mathrm{\sqrt{s_{_{NN}}}} = 200$~GeV. The solid curve (obtained with Eq.~(\ref{corrv1})) in each panel illustrates the effectiveness of the simultaneous fits, as well as the constraining power of the data.  That is, $\mathrm{v_{11}(\pT^{b})}$ evolves from purely negative to negative and positive values as the selection range for  $\mathrm{\pT^a}$ is increased.
 
The $\mathrm{v^{even}_{1}}$ extractions, were carried out for several centrality intervals at each beam energy, depending on the available statistics.  The associated systematic uncertainties were estimated from variations in the extracted values after (i) varying all of the analysis cuts by a chosen range about the standard values, (ii) crosschecks to determine the uncertainty associated with the expectation that $\mathrm{\langle \pT v^{even}_{1}(\pT) \rangle \sim 0}$ and (iii) varying the number of data points used in the fits. 
The resulting relative uncertainties, which range from $\sim 2\%$ to $\sim 10\%$, were added in quadrature to assign an overall systematic uncertainty for each measurement. The overall uncertainty for each measurement ranges from $\sim 4\%$ at $\mathrm{\sqrt{s_{_{NN}}}} = 200$~GeV and grows to $\sim 20\%$ at $\mathrm{\sqrt{s_{_{NN}}}} = 7.7$~GeV.

The resulting extracted values of $\mathrm{v^{even}_{1}(\pT)}$ for 0-10\% central Au+Au collisions are shown 
for the full span of BES-I energies in Fig.~\ref{v1_pt}. These values indicate the characteristic pattern 
of  a change from negative $\mathrm{v^{even}_{1}(\pT)}$ at low $\pT$, to positive $\mathrm{v^{even}_{1}(\pT)}$ 
for $\pT \gtrsim 1$~GeV/c, with a crossing point that only very slowly shifts with $\mathrm{\sqrt{s_{_{NN}}}}$.
This predicted pattern for rapidity-even dipolar flow \cite{Luzum:2010fb,Teaney:2010vd} 
is also indicated by the solid line in panel (a), which shows the result of a hydrodynamic model
calculation \cite{Retinskaya:2012ky}. It stems from the requirement that the net transverse 
momentum of the system is zero, i.e., $\mathrm{\langle \pT v^{even}_{1}(\pT) \rangle = 0}$, which 
implies that the hydrodynamic flow direction of low-$\pT$ particles is opposite to those 
for high-$\pT$ particles. Crosschecks made with a large sample of the data, 
confirmed that $\mathrm{\langle \pT v^{even}_{1}(\pT) \rangle \sim 0}$, within 
systematic uncertainties.
The crossing point is also expected to shift with $\mathrm{\sqrt{s_{_{NN}}}}$ since 
the $\mathrm{\langle p_{T}\rangle}$ and $\mathrm{\langle \pT^{2}\rangle}$ values change 
with $\mathrm{\sqrt{s_{_{NN}}}}$~\cite{Jia:2012gu}. For these data, there is little, if any, shift due to 
the weak dependence of the $\mathrm{\langle \pT \rangle }$ on $\mathrm{\sqrt{s_{_{NN}}}}$ for the indicated 
centrality selection. It is noteworthy that the low statistical significance of the data 
for $\mathrm{\sqrt{s_{_{NN}}}} \lt 19.6$ GeV, precluded similar centrality dependent plots for these beam energies.

The centrality dependencies of the $\mathrm{\pT}$-weighted $\lvert \mathrm{v^{even}_{1}} \rvert$ and $\mathrm{K}$  
are shown in Fig.~\ref{cent_dep} for several $\mathrm{\sqrt{s_{_{NN}}}}$ values as indicated, 
and for $\mathrm{0.4 \lt \pT \lt 0.7}$ GeV/c; this $\mathrm{\pT}$ range was selected to 
minimize the associated statistical uncertainties. The increase in the magnitude 
of $\lvert \mathrm{v^{even}_{1}} \rvert$ as collisions become more peripheral (Fig.~\ref{cent_dep}(a)), 
is expected since $\mathrm{v^{even}_{1}}$ is driven by fluctuations which become more important 
for smaller systems, i.e., for more peripheral collisions. For each value of $\mathrm{\sqrt{s_{_{NN}}}}$, 
Fig.~\ref{cent_dep}(b) indicates a linear dependence of $\mathrm{K}$ on $\mathrm{\langle N_{ch} \rangle^{-1}}$ 
with slopes that decrease with increasing $\mathrm{\sqrt{s_{_{NN}}}}$. This is to be expected since 
$\mathrm{K \propto 1/(\langle N_{ch} \rangle \langle \pT^{2}\rangle})$ and  the values for 
$\mathrm{\langle \pT^{2}\rangle}$ increase with $\mathrm{\sqrt{s_{_{NN}}}}$ for most of the centrality range.
%
%
\begin{figure}[t]
\centering{
\includegraphics[width=0.7\linewidth ,angle=-90]{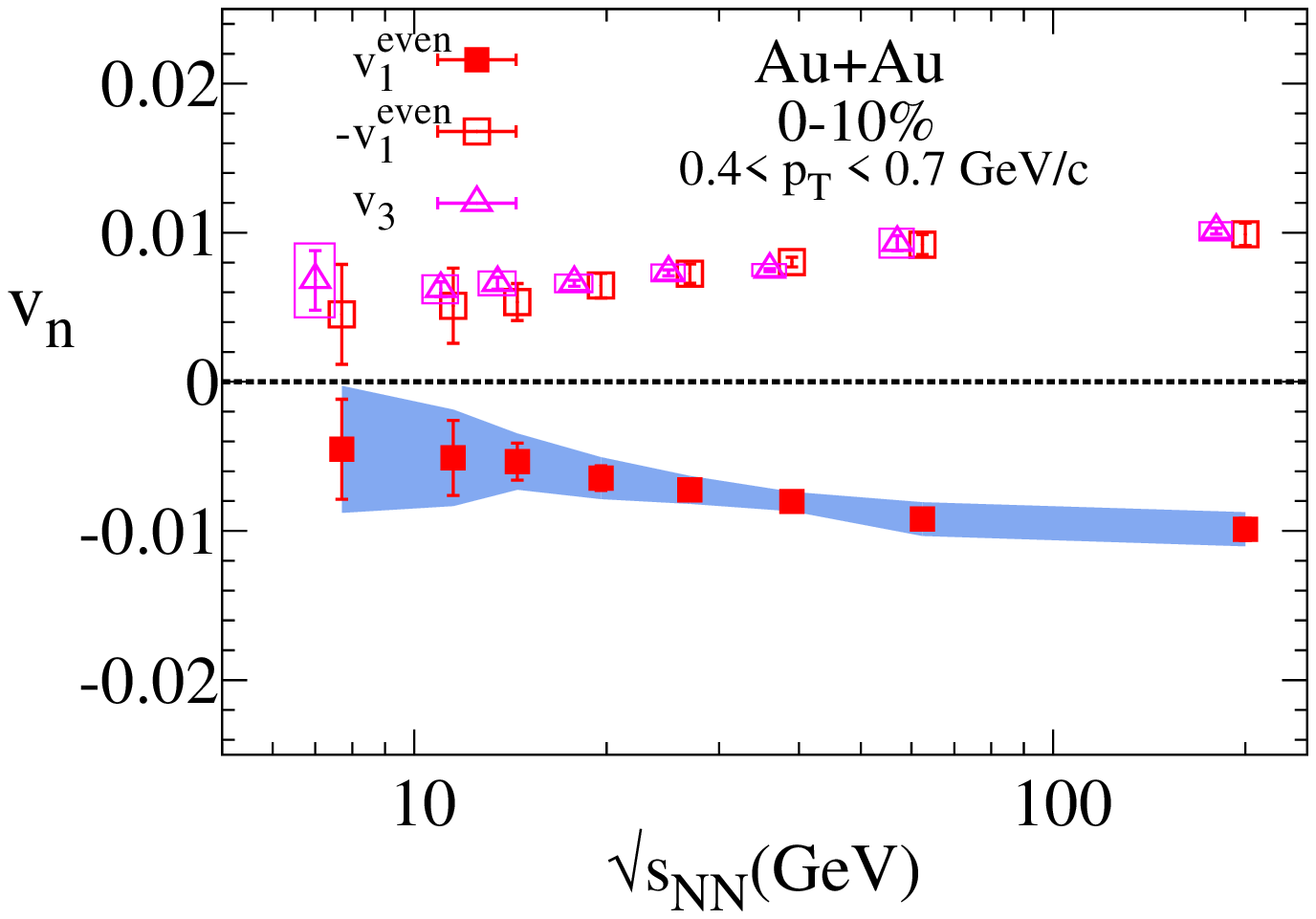}
\vskip -0.3cm
\caption{ Comparison of the $\mathrm{\sqrt{s_{_{NN}}}}$ dependence of $\mathrm{v^{even}_{1}}$  and  $\mathrm{v_3}$ 
for $0.4 \lt p_{T} \lt 0.7$~GeV/c in 0-10\% central Au+Au collisions. The $\mathrm{v^{even}_{1}}$ results 
are reflected about zero (and shifted horizontally) to facilitate a comparison of the magnitudes. 
The shaded bands indicate the systematic uncertainties.
 \label{v1_en}
 }
}
\end{figure}

Figure \ref{cent_dep}(a) also hints at both a sizable decrease in the magnitude of  
$\lvert {\mathrm{v^{even}_{1}}} \rvert$ and a possible weakening of its centrality dependence, 
as the beam energy is reduced. These patterns and the ones shown in Fig.~\ref{v1_pt} 
cannot be explained solely by the small change in the Glauber model eccentricity values 
at a given centrality which result from a change in the beam energy. Thus, they provide 
a new set of supplemental constraints for the extraction of $\mathrm{\frac{\eta}{s}(T)}$.
 
The constraining power of $\mathrm{v^{even}_{1}}$ is further illustrated in Fig.~\ref{v1_en} 
where a comparison of the excitation functions for $\mathrm{v^{even}_{1}}$ and $\mathrm{v_3}$  
is shown for $0.4 \lt \pT \lt 0.7$ GeV/c; the $\mathrm{v^{even}_{1}}$  data are reflected 
about zero to facilitate a comparison of the magnitudes. The $\mathrm{v_3}$ data, which are 
obtained from the present analysis, are in good agreement with the data reported 
in Ref.~\cite{Adamczyk:2016exq} for the same centrality and $\pT$ cuts. 
The comparison  indicates strikingly similar magnitudes and trends 
for $\lvert \mathrm{v^{even}_{1}} \rvert$ and $\mathrm{v_3}$,  suggesting a much larger 
viscous attenuation of $\mathrm{v_3}$. Note that while $\mathrm{\varepsilon_1}$ 
and $\mathrm{\varepsilon_3}$ are both fluctuations-driven, $\mathrm{\varepsilon_3 \sim 2\varepsilon_1}$ 
for 0-10\% central Au+Au collisions \cite{Teaney:2010vd,Bozek:2012hy} over the $\mathrm{\sqrt{s_{_{NN}}}}$ 
range of interest. A similar pattern was observed for comparisons made at higher $\pT$, albeit 
with lower statistical significance.  These excitation functions are expected to provide important 
experimental input to ongoing theoretical attempts to  pin down initial state models and make 
precision extractions of the specific shear viscosity.

In summary, we have employed two-particle correlation functions to carry out 
new measurements of the $\pT$ and centrality dependence of the anisotropic flow 
coefficient $\mathrm{v^{even}_{1}}$  in Au+Au collisions spanning the beam energy 
range $\mathrm{\sqrt{s_{_{NN}}}} = 7.7 - 200$~GeV. The results show the expected patterns for momentum 
conservation and the characteristic pattern of  an evolution from negative $\mathrm{v^{even}_{1}(\pT)}$ 
for $\pT \lesssim 1$~GeV/c, to positive $\mathrm{v^{even}_{1}(\pT)}$ for $\pT \gtrsim 1$~GeV/c.
That is, the trends expected when initial-state geometric fluctuations act in concert 
with hydrodynamic-like expansion to generate rapidity-even dipolar flow. 
The measured dependencies on $\mathrm{\sqrt{s_{_{NN}}}}$, centrality and $\pT$, as well as the 
similarity in magnitude and trend of the excitation functions for $\mathrm{v^{even}_{1}}$ 
and $\mathrm{v_3}$, constitute a new set of experimental constraints. 
These new constraints could prove invaluable to future theoretical attempts to 
discern between different initial-state models, as well as for precision extraction 
of the temperature dependence of the specific shear viscosity.

\section*{Acknowledgments}
%
We thank the RHIC Operations Group and RCF at BNL, the NERSC Center at LBNL, and the Open Science Grid consortium for providing resources and support. This work was supported in part by the Office of Nuclear Physics within the U.S. DOE Office of Science, the U.S. National Science Foundation, the Ministry of Education and Science of the Russian Federation, National Natural Science Foundation of China, Chinese Academy of Science, the Ministry of Science and Technology of China and the Chinese Ministry of Education, the National Research Foundation of Korea, GA and MSMT of the Czech Republic, Department of Atomic Energy and Department of Science and Technology of the Government of India; the National Science Centre of Poland, National Research Foundation, the Ministry of Science, Education and Sports of the Republic of Croatia, RosAtom of Russia and German Bundesministerium fur Bildung, Wissenschaft, Forschung and Technologie (BMBF) and the Helmholtz Association.

\section*{References}

\bibliographystyle{elsarticle-num}
\bibliography{ref_BES_v1}

\end{twocolumn}

\end{document}